\documentclass[11pt]{article}

\usepackage[dvips]{color, graphicx}

\newcounter{pubnum}
   {\begin{list}{\arabic{pubnum}}%
       {\usecounter{pubnum} \itemsep 0.1in}}
   {\end{list}}

\title{Science Icebreaker Activities:\\ An Example from Gravitational
  Wave Astronomy}

\author{Michelle B. Larson, Louis J. Rubbo, Kristina D. Zaleski and
  Shane L. Larson\\Center for Gravitational Wave Physics,
  Pennsylvania State University\\University Park, PA 16802}

\begin{document}


\maketitle


\section{Introduction}

At the beginning of a class, workshop or meeting an icebreaker activity is often
used to help loosen the group and get everyone talking.  When used as
a precursor to group learning, the icebreaker fosters communication so
later activities function more smoothly.  Science-based icebreaker activities serve the
purpose of a traditional icebreaker, while also introducing science content
to the audience. The content of the icebreaker may or may not be related to the topic of the upcoming class or meeting. Either way, the activity provides a way to get people talking while the participants simultaneously learn something new and interesting.


Several science-related icebreakers have been designed and
successfully implemented with students and adults, including a solar
activity\footnote{
\texttt{http://solar.physics.montana.edu/YPOP/Intermission/Icebreaker/}}
and a cosmic ray activity\footnote{
\texttt{http://www.chicos.caltech.edu/classroom/icebreaker/puzzle.html}}.
The subject of this article is an icebreaker activity related to
gravitational wave astronomy. Icebreakers like these could be designed 
around any subject and we encourage others to 
develop their own subject-specific science icebreakers keeping in mind two primary goals ---
help get the group interacting, and introduce some interesting science content.

Since one of the goals of this model for an icebreaker activity is to
introduce some science we begin this article by discussing a few aspects of the
developing field of gravitational wave astronomy (for the benefit of
the facilitator).  We first point
out that gravitational wave detectors do not return a pretty picture, but instead return a
time series.  We then describe the unique gravitational wave signals
from three distinct  astrophysical sources: monochromatic binaries, merging compact
objects, and extreme mass ratio encounters.  These signals form the
basis of the activity where participants work to match an ideal
gravitational wave signal with noisy detector output for each type of
source. For a more detailed introduction to gravitational wave 
astronomy and gravitational wave detectors, see {\em The Emerging Field of Gravitational Wave Astronomy} earlier in this issue.

\section{Gravitational Wave Data Analysis}

\begin{figure}[tb]
  \begin{center}
    \includegraphics[width=0.75\textwidth]{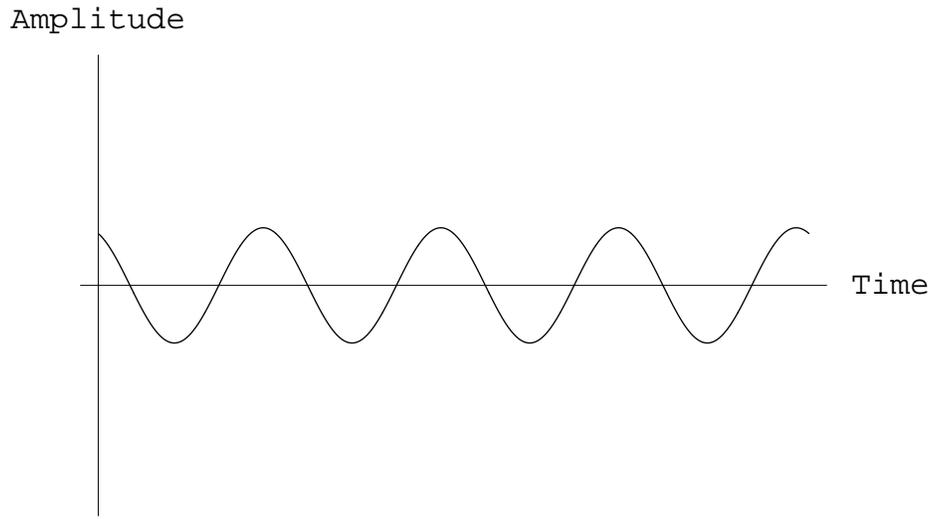}
  \end{center}
  \caption{For a binary system where the orbiting objects are far
    apart, the resulting gravitational wave signal is monochromatic.}
  \label{fig:MonoTemp}
\end{figure}

\begin{figure}[tb]
  \begin{center}
    \includegraphics[width=0.75\textwidth]{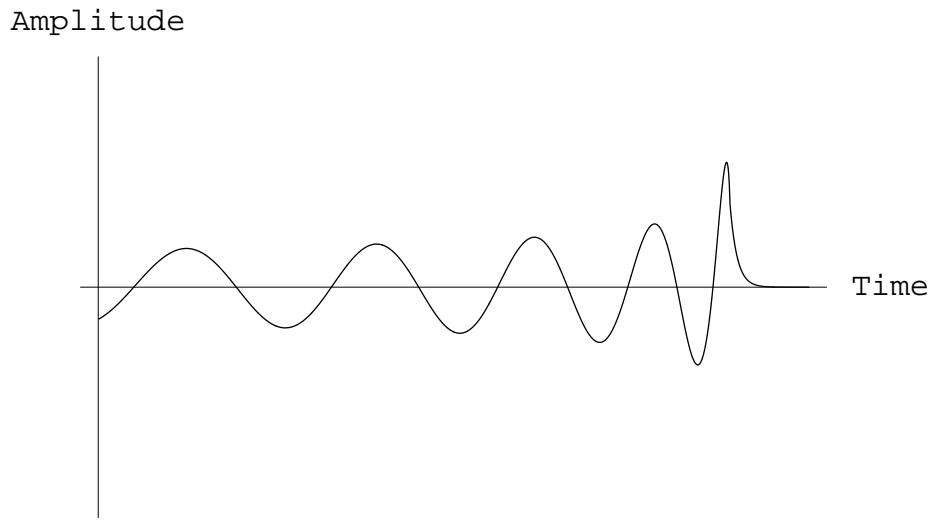}
  \end{center}
  \caption{As gravitational radiation carries energy and angular
    momentum away from the system, the binary objects slowly inspiral
    toward each other.  In the final minutes of coalescence the
    amplitudes and frequency increases rapidly, ending with the merger
    of the two objects.}
  \label{fig:CoalTemp}
\end{figure}

\begin{figure}[tb]
  \begin{center}
    \includegraphics[width=0.75\textwidth]{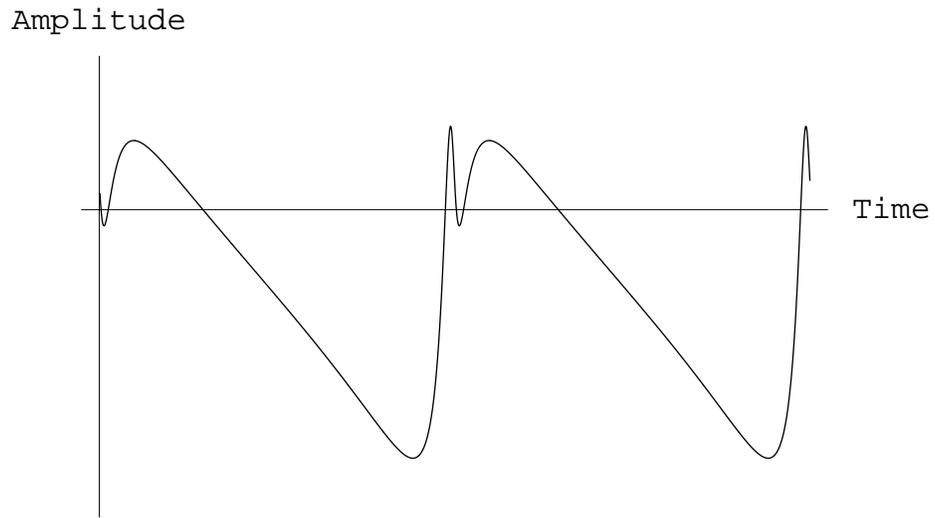}
  \end{center}
  \caption{When the ratio of the binary component masses is very
    large, the gravitational wave signal changes rapidly during the
    times when the objects are at their closest distance to each
    other.}
  \label{fig:EMRETemp}
\end{figure}

Perhaps the most challenging part of gravitational wave astronomy will
be recognizing a detection when we have one.  Gravitational wave detectors are not
imaging detectors.  They observe oscillations in spacetime, which, for
example, are made evident by measuring the relative changes in the
light travel times along different arms of an interferometric observatory.  What
will the gravitational waves from two neutron stars orbiting each
other look like?  What will two black holes colliding look like?  Or
the inspiral of a white dwarf into a super-massive black hole?
Theoretical calculations allow us to model the expected gravitational
wave signal from astrophysical systems (for example, see
Figs.~\ref{fig:MonoTemp}-\ref{fig:EMRETemp}); these models are called
{\em templates}.  Templates are the patterns gravitational wave
astronomers look for in the detector data.

As with any experiment, gravitational wave detection is complicated by
instrumental noise.  Noise can originate from a wide range of sources
such as vibrations in the Earth's crust due to ocean waves on a shore
miles away (for ground-based observatories), or minuscule changes in
the laser arm length caused by solar wind particles bouncing off the
spacecraft (for space-based observatories).  Detection involves
coaxing the signal out of the noise, and having confidence that it has
been accurately found.  Our confidence will be strengthened by
observing very strong signals, or by coincidence observations made in
several observatories at once.  Coincidence detection will eliminate
anomalous false detections due to noise in a single detector.  In this
manner, gravitational wave detectors around the world will work
together to confidently identify gravitational wave signals.

The icebreaker activity described in the next few sections models the
gravitational wave detection process described above.  Participants
identify a model signal within a noisy data stream, and compare their
solution with other groups to provide confidence that the correct
signal has been found.

\section{Astrophysical Information}

Although gravitational wave observations will not produce beautiful
images, the data tells us a lot about the source system.  At the level
of this icebreaker, participants will become familiar with the idea
that astrophysical systems can be distinguished by their gravitational
wave signals. This activity will use three different binary systems,
described below.

(1) {\em Compact star binary system}: When two massive objects orbit
each other they produce gravitational waves. If the system contains
two compact stellar objects, such as white dwarfs or neutron stars,
the gravitational radiation will be in the frequency range of current
(and future) gravitational wave detectors.  Two compact objects
orbiting in a large, circular orbit will produce a monochromatic
signal, a signal where the amplitude and frequency does not change
noticeably during the observation; Fig.~\ref{fig:MonoTemp}.

(2) { \em Coalescing binary system}: As the above binary system looses
energy through the emission of gravitational radiation, the orbit will
slowly shrink and the two objects will come closer together.  In the
final minutes the orbital period decreases rapidly and the two stars
spiral together.  During this final coalescence phase the system emits
gravitational waves with increasing amplitude and frequency, ending
with the final merger of the two objects; Fig.~\ref{fig:CoalTemp}.

(3) {\em Extreme mass ratio binary system}: Unlike the relatively
equal mass binary systems described above, the gravitational wave
signals from an extreme mass ratio system will be quite different.  An
example of an extreme mass ratio system is a compact object
(i.e. white dwarf, neutron star, or stellar mass black hole) orbiting
a super-massive black hole.  These systems are typically on very
eccentric orbits and emit a burst of gravitational waves at a time
when the smaller object is closest to the super-massive black hole.
As a result, the gravitational wave signal exhibits a rapid burst of
radiation when the two are close together; center of
Fig.~\ref{fig:EMRETemp}.

\section{Icebreaker Activity}
 
\begin{figure}[tb]
  \begin{center}
    \includegraphics[width=0.75\textwidth]{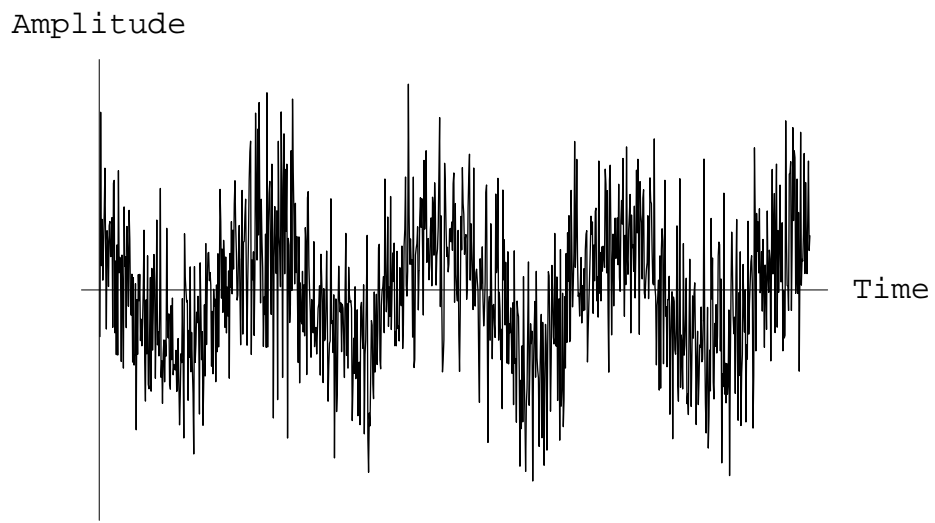}
  \end{center}
  \caption{A noisy monochromatic signal.}
  \label{fig:MonoSignal}
\end{figure}

Resources for this activity are available in printable form on the
Center for Gravitational Wave Physics' activity
page\footnote{http://cgwp.gravity.psu.edu/outreach/activities/}. The
materials allow for four activity sets; each set contains one data
stream and six templates. Each data stream contains a single source
with simulated detector noise added in.  As an example, the
monochromatic data stream is shown in Fig.~\ref{fig:MonoSignal}. The
four template files represent monochromatic and coalescing sources,
and both gravitational wave polarizations of an extreme mass ratio
source.  The six individual templates within each file have varying
signal characteristics, such as period and initial phase.  There is
also an activity key.

The icebreaker activity can be conducted with only one -- or as many as
four -- of the different activity sets. Depending on the age of your
audience, and the time available, decide in advance how many sets you
would like to include in each icebreaker packet.  Print the noisy data
signals on transparency film.  Print the gravitational wave template
files on plain paper.  All files contain two images per page.  Cut
each printed page in half creating signal and template sets of
half-sheet size for each icebreaker packet.

At the beginning of the meeting or class have everyone break into
groups of two to four people.  Pass out an icebreaker packet
(containing one to four activity sets) to each group.  Start out by
explaining that gravitational wave astronomers will make observations
of the Universe in a different way from traditional astronomers.
Explain that these are simulated gravitational wave signals from
sample astrophysical systems; a detailed description of the systems
can be saved for after the activity.  An explanation of each system is
provided in the activity key.

After the introduction, ask the groups to identify the best
template/data stream matches.  Give them ten to fifteen minutes to
complete the task.  Do not provide much instruction about what
features to use in identifying a match.  The participants, some with
encouragement, will discover the need to use the axes as reference
lines, and to look carefully at details such as amplitude, period and
initial phase differences, or other significant features in the
signals.

Encourage interaction, and allow the groups to move around the room if
they choose.  Once each group has a matched set, lead a discussion
about which template matches which data stream and why.  Facilitate
discussion about amplitude, period, and other features used when
determining their answer.  Write each group's matching solution on a
chalk board for everyone to see. If you have differences among the
groups discuss the differences and see if they can come to agreement.
If some groups matched the images based on criteria different from
what has been discussed, have them explain what they did.

Once the discussion period is over emphasize that gravitational wave
astronomers use methods very similar to these to identify sources in
real data. As part of the icebreaker activity, the discussion need not
go further than describing the different types of binary systems and
how they appear differently in the gravitational wave data.  More
detailed information can be gathered about the source system
(e.g. masses and distance to the source) by measuring the
gravitational wave signal wavelength and how the wavelength changes in
time.  A classroom extension related to these measurements will be
discussed in a follow-on paper\footnote{Rubbo, Larson, Larson and
Zaleski (in preparation)}. Finally, consider giving everyone their own
set of images (and a copy of the key) to use at home with their family
and friends!


\bigskip

\noindent This work was supported by the Center for Gravitational Wave
Physics (NSF) grant PHY-01-14375.


\pagebreak

\end{document}